# Mechanism of formation of a 2D binary alloy

L. Ottaviano,[1] G. Profeta,[1] L. Petaccia,[2] C. Nacci,[1] S. Santucci,[1] A. Pesci,[3] and M. Pedio[3]

[1] *Unità INFM and Dipartimento di Fisica Università degli Studi dell'Aquila, Via Vetoio 10, I-67010 Coppito - L'Aquila, Italy*
[2] *Laboratorio TASC-INFM, Basovizza S.S.14 Km 163.5, I-34012 Trieste, Italy*
[3] *ISM-CNR Sede distaccata di Trieste, Basovizza S.S.14 Km 163.5, I-34012 Trieste, Italy*

Direct comparison of scanning tunneling microscopy and high resolution core level photo-emission experiments provides a rationale for the mechanism of formation of a two dimensional (2D) binary alloy (1/3 mono-layer (ML) $Sn_{(1-x)}Si_x$/Si(111)-($\sqrt{3}\times\sqrt{3}$)R30°). In contrast with recent theoretical predictions, the pure metal surface ($x=0$) results partitioned into two classes (2/9 ML and 1/9 ML) of ad-atoms occupying non-equivalent $T_4$ sites. During the formation of the alloy, Si ad-atoms preferably occupy the majority type adsorption site. This peculiar substitution mechanism leads to a mutual arrangement of ad-atoms which is not random even at room temperature, but shows the typical short range order universally observed in 2D and quasi 2D binary alloys.

The physics of 2D and quasi-2D systems continuously shows exotic phenomena like high $T_C$ superconductivity [1], charge ordering [2], and low temperature phase transitions [3,4]. In particular, in view of future electronic structure engineering of 2D systems, pure two-dimensional solid solutions are of extreme interest due to the possibility to tailor their electronic properties by fine tuning the relative concentration of the two compound species. For example, the solution of semiconductor atoms into two-dimensional metals, i.e. $Pb_{(1-x)}Ge_x$/Ge(111), implies the evolution of a metal-non metal transition [5]. Similar pure 2D metallic systems, i.e. Sn/Ge(111) and Sn/Si(111), show a critical behavior at low temperatures (symmetry lowering) which is strongly influenced by a small amount (4-8 %) of extrinsic semiconductor atoms (traditionally referred to as "defects") [4,6]. There, a long range ordering of the two ad-atom species in the solid solution occurs and dictates the low temperature symmetry of the system. In order to find a rationale of the alloy structure ($x=0.5$) it is of extreme importance to follow its formation as a function of the relative semiconductor ad-atom concentration $x$, starting from a thorough understanding of the pure metal ($x=0$) surface. Scanning Tunneling Microscopy (STM) and core level photoemission spectroscopy are formidable experimental tools for such an investigation. STM furnishes a direct visualization of the mutual arrangement of the two ad-atom species of a binary alloy, and can also accurately determine the local electronic response of a lattice to an extrinsic atom [7]. On the other hand, only high-resolution photoemission spectroscopy can provide, via an accurate determination of the core level spectral line-shape, fine details that give crucial information on the different chemical environment of surface adatoms.

In this letter, the 1/3 ML $Sn_{(1-x)}Si_x$/Si(111)-($\sqrt{3}\times\sqrt{3}$)R30° solid solution, where the relative concentration of the two ad-atom species can be finely tuned [8], has been studied by STM and variable temperature (300-40 K) high resolution (40 meV) core level photo-emission experiments. In this way, at each specific $x$ value, the core level line-shape can be assigned to the corresponding STM image. This direct comparison provides a microscopic understanding of the formation mechanism of the 2D binary alloy.

The STM experimental set-up has been described in [7]. The VUV beam-line (ELETTRA Synchrotron), where the Sn 4d core level photo-emission spectra were measured with an overall energy resolution of 40 meV, is described in [9]. Samples have been prepared according to already published procedures [10,11] and we verified with STM that the post deposition annealing temperature to obtain an almost pure Sn surface of 1/3 ML Sn/Si(111)-α phase is 650 °C. Starting from this nearly perfect ($\sqrt{3}\times\sqrt{3}$)R30° surface we finely tuned the surface density of substitutional Si atoms by further annealing the samples at various temperatures up to 870 °C. Annealing at 900 °C caused complete desorption of the Sn ad-atoms. The temperature error is estimated to be within ±15 °C.

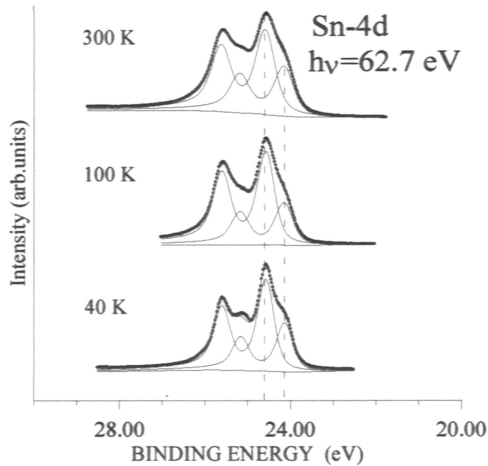

**Fig.1**
Sn 4d core level spectra of the almost defect free surface of 1/3 ML Sn/Si(111)-α phase at RT, 100 K, and 40 K. The data have been fitted with two Doniach-Sunjic doublets convoluted with a Gaussian after subtraction of an integrating background. The fitting parameter was Lorentzian width 0.14 eV, singularity index 0.08, spin-orbit split 1.04 eV, branching ratio 0.70. The Gaussian width (assumed to be the same for the two doublets) was 0.15 eV at RT, 0.11 eV at 100 K, and 0.09 eV at 40 K. The two vertical dashed lines mark the peak energies of the two components separated of 0.46 eV.

In the analysis of the Sn 4d core level spectra of the almost defect free Sn/Si(111)-($\sqrt{3}\times\sqrt{3}$)R30° surface (Fig.1), there are essentially two well-defined components: a majority one at higher binding energy is 2.0 ± 0.1 times the minority one. The spectra acquired at 100 K and 40 K, respectively, show that this partition is unchanged. There is a marked narrowing of the spectral line width of the two components which allows a more reliable fit giving as a result an almost exact intensity ratio of 2:1 between the two. Noteworthy the binding energy shift of the two components does not vary appreciably within 0.01 eV. There is no sign of any trend to recover a "single component" picture as naturally expected for a simple ($\sqrt{3}\times\sqrt{3}$)R30° surface and from the predicted lack of a soft phonon mode [12]. Uhrberg et al. [10] have presented similar findings at 70 K but our data are taken at a temperature definitely below the temperature limit (around 65 K) where any vertical Sn ad-atom vibration is theoretically expected to be frozen [12]. The picture of two non-equivalent complementary 3x3 sub-lattices (Honeycomb (HC) 2/9 ML ad-atoms, and hexagonal (HX) 1/9 ML ad-atoms) is the natural implication of this result. There are strong indications that this non-equivalence is not structural as indicated in Ref.4. It is very likely that fine electronic effect, not considered in calculations based on density functional theory in local density approximation (DFT-LDA) [12], should be taken into account. All the information accumulated so far for the *Sn/Si(111)* system at low temperature show that: i) the system remains metallic up to 70 K [10], ii) there are two Sn 4d core level components up to 40 K (this work), iii) there is no structural distortion up to 60 K [4], and to 0 K from theory [12]. Everything leads to speculate that Sn/Si(111) is at midway between the 1/3 ML Si/SiC(0001)-($\sqrt{3}\times\sqrt{3}$)R30° Mott-Hubbard insulator (gap opening and non-structural distortion [13]), and the less correlated metallic distorted [12,14] Sn/Ge(111) system.

STM direct visualization of the $Sn_{(1-x)}Si_x$/Si(111)-($\sqrt{3}\times\sqrt{3}$)R30° solid solution, obtained after post deposition annealing of the defect free surface (*x*=0), is presented in Fig.2 at various percentages of the Si ad-atoms: (a) 1 %, (b) 10 %, (c) 23 %, (d) 54 %. These surfaces have been obtained by depositing the same amount of initial Sn and after annealing for 180 s at 650 °C (a), 720 °C (b), 770 °C (c), 850 °C (d), respectively. Noteworthy what is important for the surface Si ad-atom density is only the peak temperature reached during the annealing [15]. Our estimate for an upper limit of the substitutional Si concentration, obtained annealing the samples at 870 °C, is 61%. This value is close to the limit of 1/9 ML Sn and 2/9 ML Si at the surface (corresponding STM data are not reported for brevity) and it is well above the upper limit of 1/6 ML metal ad-atom coverage reported for similar systems (Pb/Si [16], Pb/Ge[5]). Si ad-atoms are the dark spots surrounded by Sn ad-atoms with increased apparent height in filled state images [8].

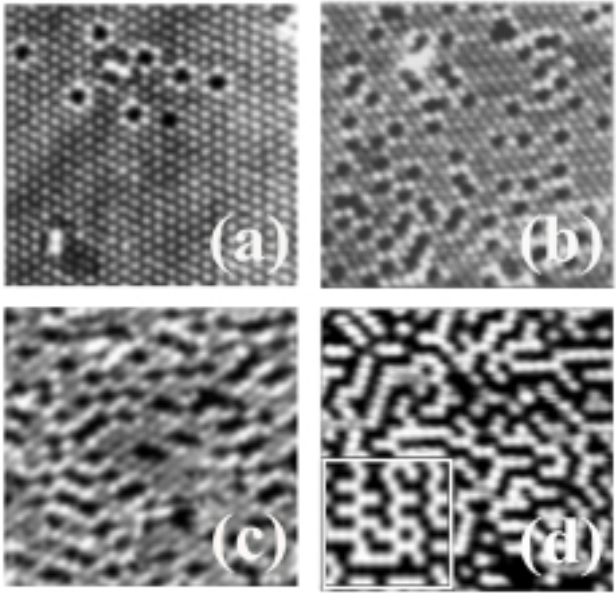

**Fig.2**
Room temperature STM filled state images (-1.5 V, 0.1 nA, 150x150 Å$^2$) of the $Sn_{(1-x)}Si_x$/Si(111) system at Si surface densities of: (a) 1 % (annealing at 650 °C), (b) 10 % (annealing at 720 °C), (c) 23 % (annealing at 770 °C), (d) 54 % (annealing at 850 °C). Lower left inset of panel (d), same length scale: result of a corresponding (46 % Sn, 54 % Si) Monte Carlo simulation of the surface.

The simple visual inspection of the STM data as a function of the Si ad-atom density already testifies a clear evidence: at intermediate Si densities (10 % and 23 %, Fig.2 panels (b) and (c), respectively) Si adatoms locally occupy second nearest neighbor sites. This corresponds to local preferential occupation of one of three complementary 3x3 sub-lattices of the surface [3]. STM thus seems to indicate that, being the occupation sites of the pure metal (x=0) surface not equivalent (Fig.1), Si ad-atoms preferentially substitute one type of Sn ad-atoms. Eventually, when the surface is almost equally shared by the two ad-atom species, the "mosaic" phase is obtained. This phase can be found ubiquitously in other pure 2D binary alloys (Pb/Si [16], Pb/Ge [5]) and also at the surface layer of bulk alloys (i.e. $Pt_{25}Co_{75}$ [17] and other similar systems referenced therein). There, a well-defined local order takes place with alternating disordered lines of Sn and Si ad-atoms (sit on an adatom, there are in the average two (four) neighbors of the same (different) species). Indeed, the local 3x3 arrangement of the Si adatoms is still the basic ingredient to fabricate the mosaic structure. Effectively, the "ad-hoc" simplest assumption of sole Si-Si nearest neighbor (NN) short range repulsion (that intrinsically arranges Si adatoms into second NN sites) yields, in a Monte Carlo simulation, a black and white ball image of the 2D alloy in striking self-similarity with the real STM appearance of the mosaic phase (see Fig.2(d) and its lower left inset).

How does this 3x3 local arrangement of Si ad-atoms, microscopically observed with STM, manifests itself in the core level spectra of Sn as a function of the Si surface concentration? Albeit eventually misleading, it might seem straightforward a tentative prediction of the expected Sn 4d core level line-shape as a function of $x$ based on the following arguments. Let us observe that, at whatever Si surface density one can identify in the STM images at least two types of Sn ad-atoms: those unperturbed (with six Sn NN) and the "bright" Sn ad-atoms, which are neighbors of Si ad-atoms. From STM it can be estimated that at 10 % (24 %, 54%) Si adatom density, "bright" Sn are 26 % (44 %, 100%) of the overall Sn ad-atoms. The assignment to these peculiar Sn ad-atoms of a well-defined spectral component (probably shifted for screening effects) in the Sn 4d core level spectra seems rather natural and was also adopted by Kidd et al. in the detailed interpretation of the Sn 4d core level spectra of Sn/Ge(111) [18]. Similarly, for the α-phase of Sn/Si(111) Uhrberg et al. have identified, on a surface showing with STM just 1 % Si ad-atom concentration, an high binding energy component (namely C3 in Fig.4 of Ref.10) accounting consistently for the 6% intensity of the overall Sn 4d peak area [10].

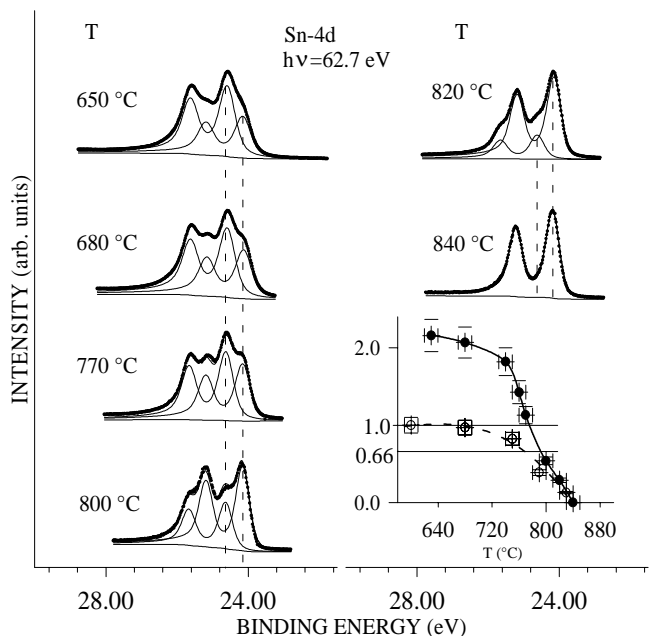

**Fig.3**
Room temperature Sn 4d core level spectra of 1/3 ML $Sn_{(1-x)}Si_x$/Si(111) obtained by annealing the system at increasing temperatures. See caption of Fig.1 for detail on the fitting procedure and on the two vertical dashed lines meaning. Lower right inset: (solid curve) relative intensity of the high-low binding energy components in the spectra, (dashed curve) absolute surface Sn amount as a function of the post deposition annealing temperature (see text for details).

Fig.3 shows our core level spectroscopy results for surfaces prepared with identical procedures to those producing the STM images of Fig.2, and gives instead clearly different results. There, the Sn 4d core level spectra are reported as a function of the post deposition annealing temperature, i.e. of the Si ad-atom density. The experimental information can be summarized as follows: i) with increasing the defect density it is not observed any marked increase of an higher binding energy component (like C3 in Ref.10) previously assigned to the Sn defect neighbors; ii) the core level spectra can always be fitted by only two components at the same binding energy values of the two characterizing the almost defect free surface; iii) the relative binding energy shift of the two components is the same in all the spectra; iv) the increase of the Si surface density is characterized in the core level spectra by a marked decrease affecting mainly the majority component at higher binding energy; v) the spectrum corresponding to the mosaic phase (T=850 °C in Fig.2) shows the complete disappearance of the higher binding energy component. Points iv) and v) are substantiated by the graphs in the inset of Fig.3. The dashed line graph shows, as a function of the post deposition annealing temperature, the absolute variation (starting from 1/3 ML of Sn) of the intensity of Sn 3d (by far the most intense peak amongst Sn typical spectra) as measured with standard XPS (Al $K_\alpha$ source equipment of Ref.7). Noteworthy, in correspondence of a 33% reduction of the integrated Sn 3d signal, the ratio of the higher to lower binding energy component in the Sn 4d spectra, as measured with synchrotron radiation (solid curve), goes to 1. Therefore, within experimental uncertainty, there is no mutual intensity rearrangement between the two components but just the disappearance of one of the two. These data hardly reconcile with the interpretation of the Sn 4d core level spectra proposing a charged ad-atom model [18]. On the other hand, there is an overall interpretation scheme that arises from these experimental data. Two types of Sn ad-atoms live onto two complementary HC and HX sub-lattices, as indicated by the data taken at 40 K on the almost defect free surface (Fig.1). The obvious HC/HX ad-atom population ratio is 2:1, but the measurements performed increasing the Si surface density show that the HC Sn ad-atoms are, in a sense, more volatile. As the density of surface substitutional Si increases this occurs only at the expense of the HC type Sn ad-atoms. Accordingly, the disappearance of the majority component in the Sn 4d core level spectra is the only important spectroscopic consequence of the increase of surface Si. Eventually, almost all (70-80%) the Sn ad-atoms of this sub-lattice can be substituted by Si atoms without changing the symmetry of the surface. This finding is completely consistent with the correspondingly measured density (and mutual correlation) of Si adatoms in the forming 2D alloy observed with STM (Fig.2). In other words, 1/3 ML (Sn-Si)/Si(111) appears to be composed by 1/9 ML tightly bound Sn ad-atoms and 2/9 ML totally miscible solution of Sn and Si. Accordingly, it follows as a natural consequence the observed atomic intermixing of the two surface species and its typical 3x3 symmetry short range order.

In conclusion, we have compared the Sn 4d photo-emission spectra and the corresponding STM images of the 1/3 ML $Sn_{(1-x)}Si_x$/Si(111) 2D solution as a function of the relative Si adatom concentration ($0<x<0.54$). On the pure metal ($x=0$) surface we have clearly and conclusively shown the existence of two non-equivalent types of $T_4$ adsorption sites in an exact ratio of 2:1. This occurrence implies the 3x3 symmetry for the ground state of this system, which is not predicted by state of art electronic structural calculation [12]. Increasing the Si ad-atom concentration, we have demonstrated that there is a selective occupation of the majority type adsorption sites only, that justifies the short range order that is microscopically measured with STM. Speculation can be made that the same type of formation mechanism produces many other 2D binary alloys [5,16,17].

Thanks are due to Roberto Cimino for helpful discussion, and to Silvio Modesti for suggestions during the preparation of the manuscript. L.P. acknowledges financial support of MURST cofin99 (Prot. 9902332155), Regione Friuli-Venezia Giulia 98, and INFM PAIS-F99.